# Three Dimensional Expansion of Margins for Single-fraction Treatments: Stereotactic Radiosurgery Brain Cases[*]

**Qinghui Zhang, Maria Chan, Yulin Song, Chandra Burman**
Department of Medical Physics, Memorial Sloan Kettering Cancer Center, New York, NY 10065
Email: zhangq@mskcc.org

**Abstract**

**Purpose**: To derive a clinically-practical margin formula between clinical target volume (CTV) and planning target volume (PTV) for single-fraction stereotactic radiosurgery (SRS). **Methods**: In previous publications on the margin between the CTV and the PTV, a Gaussian function with zero mean was assumed for the systematic error and the machine systematic error was completely ignored. In this work we adopted a Dirac delta function for the machine systematic error for a given machine with nonzero mean systematic error. Mathematical formulas for calculating the CTV-PTV margin for single-fraction SRS treatments were proposed. **Results**: Margins for single fraction treatments were derived such that the CTVs received the prescribed dose in 95% of the SRS patients. The margin defined in this study was machine specific and accounted for nonzero mean systematic error. The differences between our formulas and a previously published formula were discussed. **Conclusion**: Clinical margin formulas were proposed for determining the margin between the CTV and the PTV in SRS treatments. Previous margin's recipes, being derived specifically for conventional treatments, may be inappropriate for single-fraction SRS and could result in geometric miss of the target and even treatment failure for machines possessing of large systematic errors.

**Keywords:** Margin, cone-beam CT, image-guided radiation treatment, radiosurgery, brain tumor.

## 1. Introduction

Driven by rapid advances in on-board imaging technology, image-guided single-fraction stereotactic radiosurgery (SRS) has become increasingly popular for the treatment of both primary brain tumors and solitary brain metastases today. Its benefits and downsides have been extensively studied in the theoretical papers [1, 2]. Dictated by its fractionation scheme, a single-fraction SRS treatment has no inter-fraction setup-error distribution for a specific patient, a phenomenon that only exists in multi-fraction treatments for a specific patient. Because of its high prescription dose, one needs to make every effort to minimize the setup errors prior to the initiation of a single fraction treatment. By enhancing the geometric accuracy of radiation therapy (RT), improvements may be achieved in terms of tumor control probability, reduction in toxicity, and conformal avoidance by reduction of individualized planning target volume (PTV) margins. To minimize setup errors, historically, SRS frames have been used to replicate the initial simulation geometry at the time of therapy. Lately, medical linear accelerator (linac) manufacturers have developed integrated imaging systems to improve and facilitate the visualization of patient anatomy [4-5]. These imaging systems often use the accelerator isocenter (iso) as the iso of their system. No matter how advanced the technologies are, there are always machine systematic errors present, which are defined as the geometric misalignment between the accelerator iso and the imaging system iso. This misalignment will ultimately affect the accuracy of the replication of a simulation. This intrinsic hardware uncertainty should be incorporated into the design of the CTV-PTV margin.

While the estimation of the CTV-PTV margin has been previously studied for conventional fractionation [6-10], the methodology is inappropriate for single-fraction treatments. Specifically the systematic and random re-sidual setup errors in conventional fraction treatments are not directly applicable to single-fraction treatments. In conventional fraction treatments, the random setup error "blurs" the dose distribution isotropically, but the systematic error shifts the dose distribution unidirectionally.

Though the margin formula was derived almost twenty years ago [8], it is still the only recipe being widely implemented clinically today. However, a flaw of the derivation [8] was the zero mean assumption of the systematic and random set up errors. Owing to this, the validity of its natural extension to modern linacs with multiple rotation axes and x-ray source is being challenge. In a single-fraction treatment, both machine systematic and patient random setup errors displace the dose distribution with respect to the planned distribution. Hence, the purpose of this study is to determine the characteristics and magnitude of the uncertainty and analytically derive the CTV-PTV margin using a model-based approach. In this model, the nonzero systematic error for a specific machine is explicitly included in the CTV-PTV margin. To the best of our knowledge, this type of study has not been previously addressed in the literature.

In this paper, we assume two coordinate systems: one whose origin is affixed to the iso of the cone-beam computed tomography (CBCT) and the other one whose origin is placed at the linac's iso. For image guided SRS (IG-SRS) cases, the planning CT (PCT) iso is at the iso of the linac. Upon successful completion of CBCT-PCT image registration, one assumes that the CBCT origin coincides with that of the PCT or linac. However, due to the limitations in the alignment accuracy between the linac and CBCT isocenters, CT image quality, and human factors, there exist a number of uncertainties in the patient setup process. These uncertainties can be classified into two types: (1) systematic errors, which are mainly caused by, for example, image quality and iso accuracy limitations and (2) residual setup errors, such as 4 degrees-of-freedom (DOF) couch (three translations and one rotation) which precludes the correction of all 6 DOF errors (three translations and three rotations). It should be noted that deformation is ignored here. Therefore, residual errors always exist in the setup. Among the two, the machine systematic errors, such as iso discrepency, are machine specific and nearly patient-independent, whereas the residual setup errors vary from patient to patient. In our study, the machine systematic error refers specifically to the iso discrepancy between the linac and CBCT, which is assumed to be zero in the treatment planning process.

Systematic errors are not a new concept and have been reported in the literature [6-10]. However, there are some differences in the definition of systematic errors between this paper and previous publications. In a previous study [8], the systematic errors were designated as the "setup error at the scanner, delineation error, and motion error," which are different from the errors in iso discrepancy described in this paper. In our paper, we disregard those non-IGRT related systematic errors because they are not in the scope of our study. For example, variation among physician CTVs is irrelevant to our calculation assumptions as long as the CTV contour encompasses all the gross tumor volume and microscopic extensions of the disease. Also, the same authors assumed that their systematic errors were stochastic over a group of patients. In the paper presented here, iso differences are independent of patients and specific to machines. In another published report, the mean systematic error was described as the average for all patients in a given group. It was also assumed that one might eliminate this systematic error by "some subtraction" method (for example, couch shift) [11]. Therefore, the margin definition there did not include any information on the mean setup error. Additionally, from a technical point of view, this method is un-implementable for sub-millimeter systematic error. In this paper, we will discuss the iso differences that cannot be obtained by the average method given in previous publications. We demonstrate a simple method to include this small systematic error in the margin formula.

For a comparison, similar to a previous work [8] the margin formula for all machines and the whole population (for single-fraction treatment) is derived in this paper. Unlike the previous publications, which assumed that the "mean value" can be eliminated and consequently omitted in the margin formula. This "mean value" is included in our formulas because most IGRT systems are not capable of applying shifts at a sub-millimeter scale.

The paper is organized in the following way: In Section II (Methods), the relation between the probabilities that patients receive the prescribed dose and margins are derived for 3D expansion. Margin formulas for a group of machines and whole population are also derived here. In Section III (Results), the margin formulas for 3D expansions are given for cases in which the CTVs receive the prescribed dose in 95% of the treated patients. The differences between our margin formula and previously published one are also addressed. In Section IV, our formulas are discussed in greater detail and conclusions are given thereafter.

## 2. Methods

As mentioned earlier, in our approach, the "mean error" is included in our formula. Figure 1 explains why the margin for cases with nonzero systematic error is different from that for cases with zero systematic error. Here, AB and CD represent the CTVs in the PCT image and the corresponding verification CBCT image, respectively. Figure 1a shows that the iso of the CBCT coincides with that of the linac. Considering this, we only need to add a suitable margin around the CTV in PCT to account for the possible random setup errors, giving the PTV, indicated by EF in Figure 1a. However, if these two isocenters are not coincident (Fig. 1b), the two images are actually misaligned although they appear matched on the imaging console. As a result AB (CTV in the PCT) and CD (CTV in the CBCT) are mis-registered. To correct this mismatch, one can apply a couch shift, as has been described in a previous study [11]. However, after couch shifts, submillimeter mismatch may still remain. To include or eliminate this residual misalignment in our treatment planning process, we can adjust the CBCT iso to exactly match the radiation iso (or vice versa; green line in Fig.1b). However, submillimeter adjustments are not technically feasible. Alternatively, we can determine a proper margin to account for the subtle difference and correct for the misalignment. In Fig.1b, GH represent the PTVs, taking into account for the isocenter misalignment and residual setup error.  Even if these two isocenters are coincident, it would be still interesting to see the margin differences between the cases with and without isocenter discrepancies. $D_{CT}(\vec{x})$ represents the dose distribution in the PCT, calculated with the origin of the coordinate system at the linac isocenter.  Due to combined random setup and systematic errors, there is a displacement between the radiation and patient (CBCT) isocenters. This displacement, denoted as $\vec{V}$, is a three-dimensional (3D) vector ($\vec{V} = \vec{Iso}_{CBCT} - \vec{Iso}_{linac}$). For intracranial SRS treatment, internal organ movement is considered negligible. In the CBCT frame of reference, i.e, the patient frame of reference, the patient dose distribution can be expressed as:

$$D_{CBCT}(\vec{x}) = D_{CT}(\vec{x}+\vec{V}) \qquad (1)$$

The patient dose at point $\vec{x}$ in the CBCT image is the same as the dose in the PCT at point $\vec{x}+\vec{V}$ because of the shift in an actual patient (Fig. 1b). In other words, each point in the PTV is shifted by $\vec{V}$ from the radiation isocenter. Vector $\vec{V}$ contains two components, systematic shift ($\vec{V}_s$) and random shift ($\vec{V}_r$):

$$\vec{V} = \vec{V}_s + \vec{V}_r \quad . \qquad (2)$$

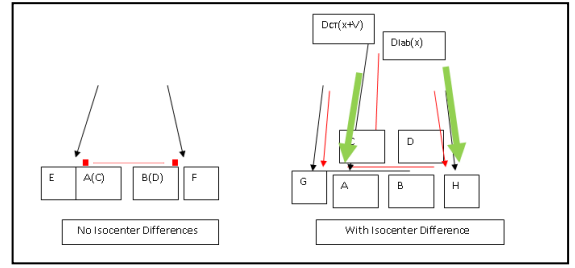

**Figures 1(a) and 1(b):  Illustration shows that the margin for cases with nonzero systematic error is different from that for cases with zero systematic error. Here, AB and CD represent the CTVs in the PCT and the corresponding CBCT, respectively. Fig. 1(a) shows that the isocenter of the CBCT is coincident with that of radiation.   EF is the PTV. Fig. 1(b) shows that these two isocenters are not coincident. The two images are actually misaligned although they appear matched on the imaging console.  Now AB and CD are misaligned.  The green lines demonstrate the case when one can shift the radiation isocenter to the CBCT isocenter.  GH is the PTV, which takes into account the misalignment between the isocenter and the residual setup error.   The patient dose is $D_{lab}(x)$. The dose for the corresponding point in the PCT is  $D_{ct}(x+V)$.**

In Eq. (1), based on convention [12-13], we make the following assumptions: (1) change in beam profiles within the PTV region is negligible; (2) CT numbers do not change dramatically; and (3) surface curvature does not change appreciably.  Assumption (1) is not appropriate for very small (<1cm) field sizes due to penumbral effects.  In clinical implementation of single-fraction SRS at our institution, 3D conformal technique is used and the prescribed 80% isodose surface covers the whole PTV. Assumption (2) is valid because tissue heterogeneity is small in the brain, thus, the heterogeneity correction is not applied in our calculation. Assumption (3) is also valid given the flat surface contour of the brain relative to the small size of the beam. However, this assumption may fail for places such as the inferior posterior head.   For a patient, we need to specify the critical dose and calculate the percent CTV coverage ($V_{D_c}$) with the following equation, where $D_c$ is the 80% prescription isodose surface in this paper.

$$V_{D_C}(\vec{V}, D_C) = \frac{\int_{CTV} H(D_{CBCT}(\vec{x}) - D_C) d\vec{x}}{\int_{CTV} d\vec{x}}$$

$$= \frac{\int_{CTV} H(D_{CBCT}(\vec{x}) - D_C) d\vec{x}}{\int_{CTV} d\vec{x}}$$

Here, the integration is performed over the points inside the CTV and H(x) is the step function that equals to 1 when x>0 and is zero otherwise. $V_{D_c}(V, D_c)$ is the cumulative volume, which is a function of $D_c$ and the iso shifts. Calculating $V_{Dc}$ for all patients, we obtain a distribution and define a threshold $V_C$, which is 100% for $V_{Dc}$. We require that the probability of patients with $V_{Dc} \geq V_c$ is greater than a threshold value $P_c$. In other words, we wish to have

$$P = \frac{1}{M} \sum_{m=1}^{M} H(V_{Dc} - V_c) \geq P_c \qquad (4)$$

where M is the number of treated patients. In the limit when M approaches infinity, Eq. (4) can be rewritten as:

$$P = \frac{1}{M} \sum_{m=1}^{M} H(V_{Dc} - V_c)$$
$$= \int H(V_{Dc} - V_c) P(\vec{V}) d\vec{V} \qquad (5)$$
$$= \int H(V_{Dc} - V_c) P(\vec{V}_s, \vec{V}_r) d\vec{V}_s d\vec{V}_r.$$

where, $P(\vec{V})$ is the probability distribution of the shift vector $\vec{V}$ for all patients and can be expressed as:

$$P(\vec{V}) = \int \delta(\vec{V} - \vec{V}_s - \vec{V}_r) P(\vec{V}_s, \vec{V}_r) d\vec{V}_s d\vec{V}_r \quad (6)$$

Here $P(\vec{V}_s, \vec{V}_r)$ is the probability distribution for the systematic error ($\vec{V}_s$) and residual setup error ($\vec{V}_r$). P can be interpreted as the probability of successful treatment. In other words, P is the probability that the patient's CTV receives the prescription dose. In Eq. (5), all patients are treated identically and have the same probability distribution. This is a good approximation for IGRT treatments. To eliminate the effects of the patient's weight or tumor size in IGRT, one can divide patients into groups according to their weights and tumor sizes. In this way, a more accurate distribution can be obtained. In this case, $P(\vec{V})$ is the probability distribution for a specific group. Because the derivation is the same, the derivation in this paper applies to both cases. To complete Eq. (5), we need two functions: $D_{CT}(x)$ and $P(\vec{V})$. For a 3D conformal treatment plan, the coverage is the objective function of our plan. That is that $D_{CT}(x) \geq D_c$ is the object function of the conformal treatment plan. The exact functional form of the dose distribution within the PTV is actually not so important for the 3D conformal treatment plan. For an IMRT plan, $D_C$ is a prescribed dose to a tumor. Therefore, the CTV-PTV margin is determined mainly by the distribution function of $P(\vec{V})$ in Eq. (5). We assume that

$$P(\vec{V}_s, \vec{V}_r) = \delta(\vec{V}_s - \vec{V}_{s0}) \; \frac{1}{(2\pi\sigma^2)^{3/2}}$$

$$\exp(-\frac{(V_{rx} - V_{r0x})^2 + (V_{ry} - V_{r0y})^2 + (V_{rz} - V_{r0z})^2}{2\sigma^2}) \quad (6)$$

where, $V_{rx}, V_{ry}, V_{rz}$ are the components of the residual setup errors in three orthogonal directions. $\vec{V}_{s0}$ is the systematic error that is fixed for a specific machine. $V_{r0x}, V_{r0y}, V_{r0z}$ are the three components of the mean residual setup error ($V_{r0}$). Here, $\vec{V}_{s0}$ represents the iso differences between CBCT and the linac. Residual setup errors are included in the following cases: (1) residual setup error after registration and shift correction (<1mm in each direction); (2) original setup error that is too small and to be corrected.

### 2.1. Three dimensional expansion margins for a single-fraction SRS case and a specific machine

In this section, we will construct a model to obtain the relationship between the probability of successful treatment and the margin for a single-fraction SRS case. In the derivation, the iso difference between the CBCT and the linac is designated as the systematic error. The derivation methodology can also be generalized to include all those systematic errors that are constant during the treatment.

We describe a method of determining the CTV-PTV margin of amount C. In a 3D uniform expansion, the computer will expand a distance C uniformly from the surface of the CTV. One can easily understand this if the shape of the CTV is spherical. If $D_c$ is defined as 80% of the prescription dose and $V_c = 100\%$, and the entire PTV receives dose $D_p(x) \geq D_c$, then $V_{Dc} \geq 100\%$ as long as $|\vec{V}| \leq C$. In other words, as long as the displacement is less than the margin, the whole CTV will receive the prescription dose. Eq. (5) becomes:

$$P = \int H(C - |\vec{V}_s + \vec{V}_s|) P(\vec{V}_s, \vec{V}_r) d\vec{V}_s d\vec{V}_r$$

$$= \int H(C - |\vec{V}_{s0} + \vec{V}_r|) \frac{1}{(2\pi\sigma^2)^{3/2}}$$

$$\exp(-\frac{(\vec{V}_r - \vec{V}_{r0})^2}{2\sigma^2}) d\vec{V}_r$$

Defining $\vec{U} = \vec{V}_r - \vec{V}_{r0}$ and $\vec{W}_0 = \vec{V}_{s0} + \vec{V}_{r0}$, the above equation can be rewritten as follows:

$$P = \int H(C - |\vec{W}_0 + \vec{U}|)$$
$$\frac{1}{(2\pi\sigma^2)^{3/2}} \exp(-\frac{U^2}{2\sigma^2}) d\vec{U} \quad (7)$$

It is clear that in Eq. (7), *P* is a function of *C* and $W_0$. Integrating Eq. (7), we obtain:

$$P = \frac{-\sigma^2 \pi}{(2\pi\sigma^2)^{3/2} W_0} [-2\sigma^2 \exp(-\frac{(C+W_0)^2}{2\sigma^2})$$
$$- 2W_0 \int_0^{W_0+C} \exp(\frac{-U^2}{2\sigma^2}) dU] \quad (8)$$
$$+ \frac{\sigma^2 \pi}{(2\pi\sigma^2)^{3/2} W_0} [\exp(\frac{-(C-W_0)^2}{2\sigma^2})][-2\sigma^2]$$
$$- 2W_0 \int_0^{W_0-C} \exp(\frac{-U^2}{2\sigma^2}) dU]$$

The Solution of the above equation yields the corresponding *C* for a fixed *P*. However, the nonlinear nature makes it difficult to obtain an analytical solution. (8)

### 2.2. Three-dimensional expansion for all machines and all patients.

For a comparison, we will also calculate the margin for the case for a group of machines and all patients (single fraction). In this scenario, the systematic error is different for different machines and we assume that it is a Gaussian function, as in previous publications. We will determine the margins for the case of 3D expansion.

For all patients and all machines, Eq. (6) can be expressed as

$$P(\vec{V}_s, \vec{V}_r) = \frac{1}{(2\pi\Pi^2)^{3/2}}$$

$$\exp(-\frac{(V_{sx} - \overline{V}_{s0x})^2 + (V_{sy} - \overline{V}_{s0y})^2 + (V_{sz} - \overline{V}_{s0z})^2}{2\Pi^2})$$

$$\frac{1}{(2\pi\overline{\sigma}^2)^{3/2}} \quad (9)$$

$$\exp(-\frac{(V_{rx} - \overline{V}_{r0x})^2 + (V_{ry} - \overline{V}_{r0y})^2 + (V_{rz} - \overline{V}_{r0z})^2}{2\overline{\sigma}^2})$$

where, $\vec{\overline{V}}_{s0} = (\overline{V}_{s0x}, \overline{V}_{s0y}, \overline{V}_{s0z})$ is the average systematic error; $\overline{V}_r = (\overline{V}_{rx}, \overline{V}_{ry}, \overline{V}_{rz})$ is the average setup error; and $\overline{\sigma}$ is the standard deviation for the residual setup error for all machines. If the setup procedures are the same, we can then assume that $\overline{\sigma} = \sigma$. Bringing Eq. (9) into Eq. (5), we have

$$P = \frac{-\beta^2 \pi}{(2\pi\beta^2)^{3/2} \overline{W}_0}$$

$$[-2\beta^2 \exp(-\frac{(C+\overline{W}_0)^2}{2\beta^2}) - 2W_0 \int_0^{\overline{W}_0+C} \exp(\frac{-U^2}{2\beta^2}) dU]$$

$$+ \frac{\beta^2 \pi}{(2\pi\beta^2)^{3/2} \overline{W}_0} [\exp(\frac{-(C-\overline{W}_0)^2}{2\beta^2})][-2\beta^2]$$

$$- 2W_0 \int_0^{\overline{W}_0-C} \exp(\frac{-U^2}{2\beta^2}) dU] \quad (10)$$

where, $\beta^2 = \Pi^2 + \overline{\sigma}^2$. The similarity between Eq. (10) and Eq. (8) is very clear. Here, $\overline{W}_0 = \overline{V}_{s0} + \overline{V}_{r0}$, which is assumed to be zero in the previous margin formula.

### 3. Results

At this point, we have derived the relation between the margins and the probability that the CTV receives the prescribed dose for cases of 3D uniform expansion. The following procedures are used in the derivation of margin formula for both 3D expansions:

(1): For 3D expansion, Eq. (8) is used to obtain the relation between C and $\sigma$ for a fixed $W_0$. Polynomial functions are used to fit those relations and the corresponding coefficients are obtained.

(2) Repeating the above process for different $W_0$, those coefficients given in (1) as a function of $W_0$ can also be obtained. Because of the similarity between Eq. (10) and Eq. (8), the derivation procedure is virtually identical to the case for a specific machine.

### 3.1. 3D Uniform Expansion

In Figure 2, the probability of successful treatment as a function of margin is given for the following groups of values:
$W_0 = 1.5mm$, $\sigma = 1.0mm$; $W_0 = 1.5mm$, $\sigma = 0.5mm$; $W_0 = 1.5mm$, $\sigma = 0.1mm$; $W_0 = 0.8mm$, $\sigma = 1.0mm$. It is clear that when $\sigma$ becomes smaller, the margin C, which yields the probability of success around 100%, is closer to the systematic error (1.5mm in this example). Intuitively, the smaller the systematic error is ($W_0 = 0.8mm < W_0 = 1.5mm$), the smaller the margin becomes for the same probability of success, as indicated by the red dashed line and the blue solid line in Figure 2.

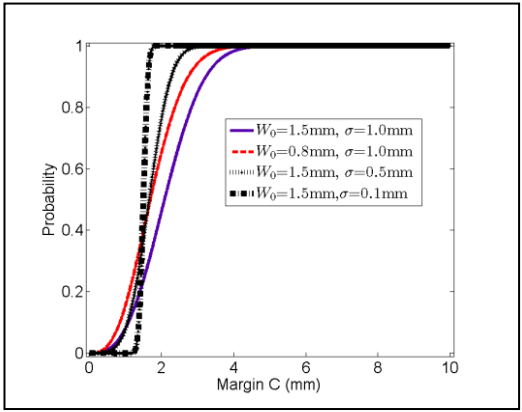

**Figure 2: Treatment success probability as a function of margin for different combinations of systematic and setup errors. Blue solid line: $W_0 = 1.5mm$, $\sigma = 1mm$; black solid line: $W_0 = 1.5mm$, $\sigma = 0.5mm$; black dot-dashed line: $W_0 = 1.5mm$, $\sigma = 0.1mm$; red dashed line: $W_0 = 0.8mm$, $\sigma = 1mm$.**

Understandably, if there is no residual setup error, then the margin C approaches $W_0$. Therefore, the general margin can be expressed (up to the second order in $\sigma$) as:

$$C = W_0 + a_0(W_0) + a_1(W_0)\sigma + a_2(W_0)\sigma^2 \quad (11)$$

where, $a_i(W_0)(i=0,1,2)$ is a function of $W_0$. Note that $a_0 = 0$ because $C \to W_0$, when $\sigma \to 0$. Hence,

$$C = W_0 + a_1(W_0)\sigma + a_2(W_0)\sigma^2 \quad (12)$$

If we accept $P = 95\%$, a plot of $C$ vs. sigma is given in Figure 3. The specific margin function are given as:

$$C = W_0 + 1.894\sigma + 0.1923\sigma^2 \ when\ W_0 = 1.5mm$$

$$C = W_0 + 2.092\sigma + 0.1835\sigma^2 \ when\ W_0 = 0.8mm$$

$$C = W_0 + 2.655\sigma + 0.0459\sigma^2 \ when\ W_0 = 0.1mm$$

Figure 3 reveals the effects of machine systematic error on the CTV-PTV margin, namely, the larger the $W_0$, the larger the margin.

We repeat the above process for several additional $W_0$ values. In other words, we first calculate $C$ vs. $\sigma$ for a different $W_0$, then we use Eq. (12) to fit the curve to get $a_1(W_0)$ and $a_2(W_0)$. Figures 4 and 5 show the behavior of $a_1(W_0)$ and $a_2(W_0)$ as a function of $W_0$, respectively. We have found that

$$a_1(W_0) = 2.787 - 1.424W_0 + 0.848W_0^2 - 0.198W_0^3 \quad (13)$$

and

$$a_2(W_0) = 0.470W_0 - 0.379W_0^2 + 0.101W_0^3 \quad (14)$$

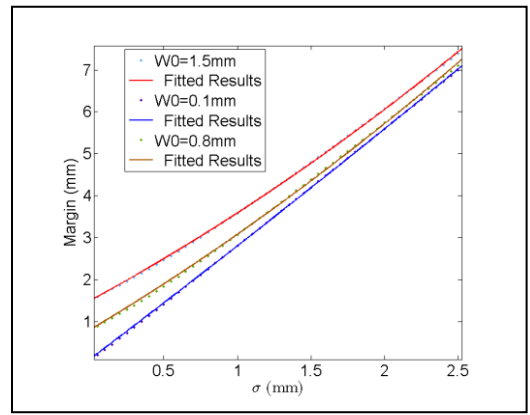

**Figure 3: Probability margin of 95% success (i.e., the CTV receives the prescribed dose) as a function of standard deviation for different systematic errors ($W_0 = 1.5mm, 0.1mm, 0.8mm$).**

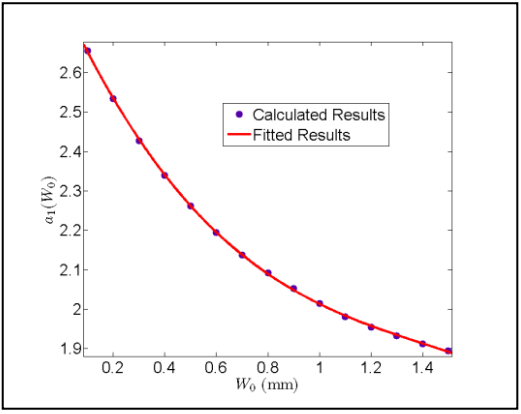

**Figure 4: Behavior of the margin parameter $a_1(W_0)$ as a function of $W_0$**

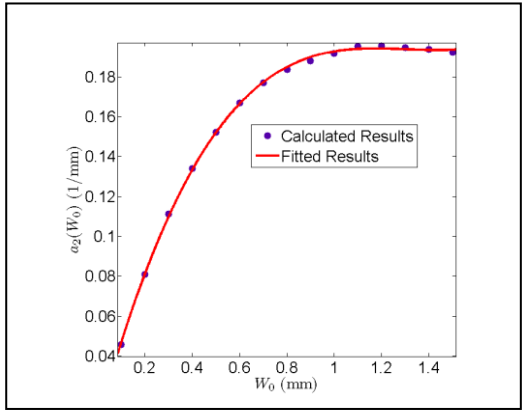

**Figure 5: Behavior of the margin parameter $a_2(W_0)$ as a function of $W_0$**

### 3.2. Margins for all patients and all machines.

The derivation of margins for the cases of all patients and all machines is quite similar to that of margins for the case of one specific machine

For a group of machines and the whole population, the margin for a single-fraction treatment is

$$C_{group} = \overline{W_0} + a_1(\overline{W_0})\beta + a_2(\overline{W_0})\beta^2 \qquad (15)$$

where, $a_1(\overline{W_0})$ and $a_2(\overline{W_0})$ have the same analytical expressions as Eq. (13) and Eq. (14). If one assumes $\overline{V}_{s0} = 0$ and $\overline{V}_{r0} = 0$, Eq. (15) is reduced to:

$$C_{group} = 2.787\beta \qquad (16)$$

Here, we need to point out that $\overline{V}_{s0} = 0$ is a very strong assumption that is not correct for a cancer center that has several IGRT machines. Treatment plan created by using Eq. (16) could seriously compromise the treatment outcome due to incomplete coverage of the CTV.

For $\Pi = 0.5mm$, $W_0 = 0.5mm$, and $W_0 = 1.2mm$, the corresponding margins calculated by Eq. (16) and Eq. (12) are given in Figure 6. It is clear that for a specific machine for which $W_0 = 0.5mm$, the patient may have undergone effective treatment but with extra radiation to normal tissue because of the overly larger margin calculated from previous margin equations than needed. However, patients treated using a machine for which $W_0 = 1.2mm$ may not receive a sufficient dose for an effective treatment because of the smaller margins used.

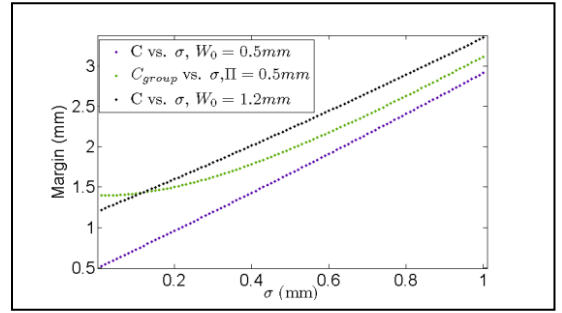

**Figure 6: The margin differences between cases of a specific machine (blue and black) and for all machines (green). When using the margin for all machines, patients could be either underdosed or overdosed, depending on the specific machine used.**

### 4. Conclusions and Discussions

CBCT technology not only increases the accuracy of radiotherapy by verifying the patient position immediately before irradiation, but also uses the enhanced or confirmed geometric accuracy to review and potentially reduce setup margins for the design of the PTV. It ultimately leads to a reduction in doses to organs at risk (OAR) and prevents potential detrimental dose escalations to these organs. However, it is also well known that the CBCT imaging iso center is not aligned with the radiation center perfectly, therefore an extra margin is needed for the CTV-PTV margin for IGRT

8                                                                                                                                          Q. H. Zhang *ET  AL*.

cases. This extra margin will become important and crucial when the random setup margin is very small.

Defining suitable CTV-PTV margins is one of the most important tasks for medical physicists in radiation treatment planning. The formula reported in previous publications [8,9] has been widely used in this field for margin determination. However, the methodology adopted to develop this formula is not appropriate for single-fraction treatments for the following reasons:

(1) One of the assumptions used in deriving this formula was that the average systematic error was zero. This might be true when one takes an average of all machines, or in a hypothetical experiment consisting of many fractions in which a single machine exhibits an error for each fraction with a mean value of zero. We believe that neither of these assumptions is applicable to single-fraction treatment [16].

(2) In a previous study [8], the margin consisted of two terms: one was the margin for the systematic error and the other was the margin for the random setup error. In our study, the machine systematic error is a delta function; therefore, the derivation procedure in the previous studies cannot be directly applied to our cases.

(3) For the margin for multi-fraction treatment, it may be acceptable for part of the CTV to be outside of the radiation field for some fractions because of its multifractional nature. However, this is unacceptable for single-fraction cases.   Finally, as we pointed out earlier, the systematic error cannot be eliminated entirely using a couch correction approach because the minimum possible shift from the imaging console is 1mm.

The margin definition described in this paper is machine-specific and more appropriate for a single-fraction treatment. This three dimensional symmetric expansion includes the nonzero systematical error. The formulas for symmetric expansion margins are given in Eq. (15) and Eq. (12). However, we should point out that for convenience, polynomial functions are used to fit the coefficients in Eq. (15) and Eq. (12). Other types of functions can also be used to fit our data.

Different from previous margin formula that was derived for multifraction treatment and a group of machines, margin formulas for a specific machine have been proposed here for determining appropriate CTV-PTV margins for SRS cases. In addition, margin formulas for single fraction and a group of machines are also derived. It has been found that this nonzero machine systematic error makes the margin formula more complex than the previous margin formula.  Although in this paper, we have concentrated only on one type of systematic errors, this methodology can be easily extended to cases with multiple unchangeable systematic errors. Nonzero systematic error is explicitly included in our margin formulas that have never been reported before.  Our derivation eliminates the assumption used in previous derivations of margin formulas that the mean systematic error is zero, therefore, it  is more general for clinical application.

We like to conclude our paper with following comments: This derivation is primarily of academic interest only when the required composited margin is large. Under this circumstance, one can either ignore this iso difference or add it to the originally calculated margin. However this approach might be an improper way to create margin if the composite margin is the same order of the iso differences. More importantly we should emphasize here that the calculation method for multi-fraction treatments cannot be applied to single fraction treatment, for which, one needs to follow the formula given in this paper. We will leave the three one dimension expansions and asymmetry expansion of CTV-PTV margin to our follow up paper [18].

## Acknowledgements

We would like to thank Drs. Howard Amols, Gig Mageras, Ellen Yorke, and Clifton C. Ling for helpful discussions.  In addition, we are grateful to Sandhya George for editing the draft   manuscript.

## REFERENCES

[1] C.C Ling, Y. C. Lo and D. A. Larson, "Radiobiophysical aspects of stereotaxic radiation treatment of central nervous system diseases," *Semin Radiat. Oncol.,*Vol. 5, 1995, pp. 192-196.

[2] Y. Lo, C. Ling and D. Larson, "The effect of setup uncertainties on the radiobiological advantage of fractionation in stereotaxic radiotherapy*,"  Int. J. Radiat. Oncol. Biol. Phys.* Vol. 34, 1996, pp 1113-1119.

[3] C. A. McBain, A. M. Henry, J. Sykes, A. AmerLo, T. Marchant, et al., "X-ray volumetric imaging in image-guided radiotherapy: the new standard in on-treatment imaging," *Int. J. Radiat. Oncol. Biol. Phys.* Vol. 64, 2006 , pp 625-634.

[4] A. Fukuda, "Pretreatment setup verification by cone beam CT in stereotactic radiosurgery: phantom study," *J.  Appl. Clin. Med. Phys.* Vol. 11, 2010 , pp 3162.

[5] J. Zhu, "Feasibility of using cone-beam CT to verify and

reposition the optically guided target localization of linear accelerator based stereotactic radiosurgery," *Med. Phys.* Vol. 38, 2011 , pp 390-396.

[6] G.S. Mageras, Z. Fuks, S.A. Leibel, C.C. Ling, M.J. Zelefsky, *et. al.,,* "Computerized design of target margins for treatment uncertainties in conformal radiotherapy," *Int. J. Radiat. Oncol. Biol. Phys.* Vol. 43, 1999, pp 437-445.

[7] J.C. Stroom, H.C. de Boer, H. Huizenga, and A.G. Visser," Inclusion of geometrical uncertainties in radiotherapy treatment planning by means of coverage probability," *Int. J. Radiat. Oncol. Biol. Phys.* Vol. 43, 1999, pp 905-919.

[8] M. van Herk, P. Remeijer, C. Rasch, J. Lebesque.C. Stroom," dose-population histograms for deriving treatment margins in radiotherapy," *Int. J. Radiat. Oncol. Biol. Phys.* Vol. 47, 2000, pp 1121-1135.

[9] D. Yan, D. Lockman, A. Martinez, J. Wong, D. Brabbins, F. Vicini, J. Liang, and L. Kestin," Computed tomography guided management of interfractional patient variation," *Semin. Radiat. Oncol.* Vol. 15, 2000, pp 168-179.

[10] D. Yan, J. Wong, F. Vicini, J. Michalski, C. Pan, A. Frazier, E. Horwitz, and A. Martinez," Adaptive modification of treatment planning to minimize the deleterious effects of treatment setup errors," *Int. J. Radiat. Oncol. Biol. Phys.* Vol. 38, 2000, pp 197-206.

[11] O.A. Zeidan, K.M. Langen, S.L. Meeks, R.R. Manon, T.H. Wagner, T.R. Willoughby, D.W. Jenkins, and P.A. Kupelian,"Evaluation of image-guidance protocols in the treatment of head and neck cancers," *Int. J. Radiat. Oncol. Biol. Phys.* Vol. 67, 2007, pp 670-677.

[12] T. Craig, J. Battista, and J. Van Dyk," Limitations of a convolution method for modeling geometric uncertainties in radiation therapy. I. The effects of shift invariance," *Med. Phys.* Vol. 30, 2003, pp 2001-2011.

[13] T. Craig, J. Battista, and J. Van Dyk," Limitations of a convolution method for modeling geometric uncertainties in radiation therapy. II. The effect of a finite number of fractions," *Med. Phys.* Vol. 30, 2003, pp 2012-2020.

[14] M.G. Witte, J. van der Geer, C. Schneider, J.V. Lebesque, and M. van Herk, "The effects of target size and tissue density on the minimum margin required for random errors," *Med. Phys.* Vol. 31, 2004, pp 3068-3079.

[15] S.F. Zavgorodni, "Treatment planning algorithm corrections accounting for random setup uncertainties in fractionated stereotactic radiotherapy," *Med. Phys.* Vol. 27, 2000, pp 685-690.

[16] W. Du, J.N. Yang, E.L. Chang, D. Luo, M.F. McAleer, A. Shiu, and M.K. Martel, "quality assurance procedure to evaluate cone-beam CT image center congruence with the radiation isocenter of a linear accelerator," *J. Appl. Clin. Med. Phys.* Vol. 11, 2010, pp 3297-690.

[17] L. Masi, F. Casamassima, C. Polli, C. Menichelli, I. Bonucci, and C. Cavedon,, ," Cone beam CT image guidance for intracranial stereotactic treatments: comparison with a frame guided set-up," *Int. J. Radiat. Oncol. Biol. Phys.* Vol. 71, 2008, pp 926-933.

[18] Q. Zhang et al," Three one dimensional expansion of margins for single fraction treatments: Stereotactic Radiosurgery Brain Cases," In preparation.